\documentclass[aps,prb,preprint]{revtex4}

\input{tcilatex}

\begin{document}

\title{Time-reversal symmetry violation and the structure of Superconducting
Order Parameter of PrOs$_{4}$Sb$_{12}$}
\author{V.G. Yarzhemsky and V.I.Nefedov}
\affiliation{Institute of General and Inorganic Chenistry of RAS,e-mail vgyar@igic.ras.ru}

\begin{abstract}
The antisymmetrised two-electron functions are constructed for the point
group T$_{h}$, i.e. the symmetry group of unconventional superconductor PrOs$%
_{4}$Sb$_{14}$., and its subgroup D$_{2h}$.The nodal structure of these
function depending on the wavevector group is investigated. Theoretical
nodal structure of these functions made possible to explain the experimental
nodal structure PrOs$_{4}$Sb$_{14}$. as a result of time-reversal symmetry
violation.
\end{abstract}

\maketitle

\section{Introduction}

The superconducting transition is accompanied by different types of symmetry
violations. \ In particular these symmetry violations \ appears as
anisotropic structure of the superconducting order parameter ( SOP),
vanishes at lines and points of Fermi surface [1]. The SOP in many of
unconventional superconductors as antisymmetrical $(UTh)Be_{13}$ [2], $%
Sr_{2}RuO_{4}$ [3,4] and $PrOs_{4}Sb_{12}$ [5-9] is triplet whose spatial
part is odd. The muon spin relaxation experiments in the three above
material magnetic fields, which result in time-reversal symmetry
violation[10].

Thermal conductivity and specific heat measurements in magnetic field,
indicated that SOP of $PrOs_{4}Sb_{12}$depends on the magnitude of the field
[11] and it was concluded that at high fields SOP has six point nodes in the
directions $[\pm 1,0,0]$, $[0,\pm 1,0]$ and $[0,0,\pm 1]$ and at low fields
SOP has two point nodes in the directions $[0,0$,$\pm 1]$. Hence it follows
that $PrOs_{4}Sb_{12}$ is the first heavy-fermion superconductor having two
phases with point nodes and a transition between them occurs in
superconducting state. It was suggested that superconductivity in $%
PrOs_{4}Sb_{12}$ is connected with electric quadrupole interaction in
distinction to other heavy-fermion superconductors.

The nodal structure of SOP is closely connected with interactions
responsible for Cooper pairing e.g. magnetic interactions instead of
electron-phonon interactions. Experimental investigations of SOP nodal
structure and theoretical interpretations of these experiments are basic to
an understanding of the nature of such interactions.  There are three group
theoretical approaches to the SOP investigations based on point groups,
continuous groups and space groups. The point-group approach to SOP
[1,12,13] is based on the reduction of spherical functions, corresponding to
isotropic case onto the crystal point group. The choice of basis functions
of irreducible representation (IR) of point group is not unique and hence
the conclusions of this approach on the nodal structure of SOP are ambiguous
[14]. According to the results of point group approach there are no symmetry
reasons for vanishing of triplet SOP on the lines of Fermi surface [13]
(Blount theorem). This statement is connected with additional degree of
freedom of spin part of the wavefunction. In this situation the majority of
triplet superconductors, in which SOP vanishes on lines of Fermi surface ,
belong to the exclusion of this theorem. We will show below that in order to
explain these exclusions, the symmetry reduction of spin part , connected
with time-reversal symmetry violations should be taken into account.

The continuous group approach unites singlet SOP and antiferromagnetic order
parameter by the operators of $SO(5)$ group [15]. In this approach a
two-fermion state is described by the five dimensional superspin with
following components: singlet order parameter for particles, singlet order
parameter for holes and three components of antiferromagnetic order
parameter.

The space-group approach to SOP [16-23] is based on the Ginzburg-Landau [24]
statement that the symmetries of SOP and of the Cooper pair wavefunction are
identical, on the Anderson symmetry description of singlet and triplet pairs
in a general point in k- space and on the induced representation method
[26,27]. In the present work making use of the space-group approach general
triplet two electron wave-functions are constructed for group $T_{h}$ and
its subgroup $D_{2h}$. The latter group corresponds to the symmetry of
triplet spin in the case of time-reversal symmetry vilation. This resultes
are used to identify the symmerty of phases with different nodal structure
of $PrOs_{4}Sb_{12}$.

\section{Method of calculations and results}

In spherically symmetric case two electrons with opposite momenta are
connected into a Cooper pair. According to Anderson [25], the wavefunctions
of electrons in singlet pair are connected by time reversal $\theta $,and
three components of triplet pair are written in second quantization form as:

\begin{equation}
\ \langle C_{k}^{\dagger }C_{Ik}^{\dagger }\rangle  \label{and1}
\end{equation}

\begin{equation}
\ \langle C_{k}^{\dagger }C_{\theta k}^{\dagger }-C_{Ik}^{\dagger
}C_{I\theta k}^{\dagger }\rangle  \label{and2}
\end{equation}

\begin{equation}
\ \langle C_{\theta k}^{\dagger }C_{I\theta k}^{\dagger }\rangle
\label{and3}
\end{equation}

where $I$ is the space inversion. In L-S coupling singlet and the three
components of triplet two-electron functions are written as:

\begin{equation}
\Psi ^{s}=\Phi ^{s}S^{0}  \label{singl}
\end{equation}

\begin{equation}
\Psi ^{t}=\Phi ^{t}S_{m}^{1}, m=-1,0,1  \label{tripl}
\end{equation}

where $S^{0}$ and $S_{m}^{1}$ are singlet and triplet functions.
Corresponding spatial parts are given by two following formulae:

\begin{equation}
\Phi ^{s}\ \ =\varphi _{1}^{1}\varphi _{I}^{2}+\varphi _{1}^{2}\varphi
_{I}^{1}  \label{singls}
\end{equation}

\begin{equation}
\Phi ^{t}\ =\varphi _{1}^{1}\varphi _{I}^{2}-\varphi _{1}^{2}\varphi _{I}^{1}
\label{tr2}
\end{equation}

where subscripts $1$ and $I$ correspond to an arbitrary chosen wave vector
in a Brillouin zone and the result of action of space inversion on it, and
superscripts denote the number of electron coordinate. Owing to translation
symmetry one-electron functions in crystal are characterized by the group $H$
of the wave vector $\vec{k}$ and the star of the wave vector $\{\vec{k}\}$.
Correct translation invariant two-electron wavefunction in crystal is
expressed as antisymmetrical linear combination of the wavefunction
belonging to all prongs of the star. It follows from the induced
representation theory, [21,26,27] that these functions may be constructed by
the action of projection operators on the functions (4) and (5). It follows
from the reciprocity theorem [27], that the projections are non zero for
those IRs of the whole group , whose characters are not orthogonal to
characters of IRs of the subgroup $H+IH$, corresponding to function (4) and
(5). Hence it follows that for $\vec{k}$ a general point in a Brillouin zone
all even IRs are possible for singlet pair and all add IRs are possible for
triplet pair and the number of appearance of each IR equals to its
dimension. If the wave vector group$H$ consists of more then one element,
the characters of IRs for singlet and triplet pairs \ are given by two
following formulae:

\bigskip

\begin{equation}
\ \chi ^{\pm }(h)=\chi ^{2}(D(h))  \label{x1}
\end{equation}

\bigskip $\ $

\begin{equation}
\ \chi ^{\pm }(\delta h)=\chi (D(\delta h\delta h))  \label{x2}
\end{equation}

where $h$ is an element of $H$, $D$ is IR of $H$, the plus sign corresponds
to singlet pair and minus sign corresponds to triplet pair and $\delta $ is
an element of double coset in decomposition $G$ with respect to $H$. Since
the total momentum of a Cooper pair equals zero, we consider the double
cosets defined by $\delta $ reversing the direction of $\vec{k}$. For k a
general point of a Brillouin zone $\delta $ is a space inversion only and in
symmetrical direction we can also take as $\delta $ a 180$^{o}$ rotation
around the axis perpendicular to $\vec{k}$.

For real one-dimensional IRs of H we obtain from formulae (8) and (9) the
unit representation of $H+\delta H$ for singlet pairs. For triplet \ pairs
characters of all elements unchanging $\vec{k}$ equal to 1 and characters of
all elements changing direction of $\vec{k}$ equal to -1. It is clear that
in general case this two IRs don't exhaust all IRs of $H+\delta H$. \ It can
be easily show that if IRs of H are two-dimensional, there are singlet pairs
with odd spatial part and triplet pairs with even spatial part [16]. Thus
formulae \ (8) and (9) generalize Anderson approach for the case of
symmetrical directions and planes in a Brillouin zone.

Let us consider the case of $\vec{k}$ on the plane of symmetry, i.e. $%
H=C_{s} $. We easily obtain that spatial part of singlet pair belongs to IR $%
A_{g}$ of $C_{2h}$ and that of triplet pair belongs to IR $B_{u}$ of $C_{2h}$%
. In the case of singlet pairs the absence of another even IR $B_{g}$ \
implies, that some of the even IRs of the whole group will be forbidden for
two-electron states on this plane. For triplet pairs without spin the
absence of another even IR also implies that some odd IRs will be forbidden
at this plane. But the account for the spin lifts this restriction. In the
case of time-reversal symmetry one can take real triplet spin basis
functions, which are usually denoted as $\hat{x}$, $\hat{y}$ and $\hat{z}$.
On the symmetry planes these functions belong to two different even IRs of
group $C_{2h}$\ and their products on the IR of spatial part $B_{u}$ include
all IRs of $C_{2h}$. \ Hence it follows that there are no symmetry
requirement for vanishing of triplet SOP on the planes of symmetry (Blount
theorem[13]). \ The existence of internal magnetic field results in
time-reversal symmetry violation. In this case different directions of spin
become non-equivalent and it is natural to suppose that only one of them
corresponds to Cooper pairing. In this case the number of possible IRs on
the planes of symmetry is reduced and symmetry restrictions fro IRs of the
whole group become possible. Let one of IRs, forbidden on the plane of
symmetry corresponds to Cooper pairing. The intersection of this plane with
the Fermi surface results line on nodes. If the IR, corresponding to Cooper
pairing forbidden on the line of symmetry, the intersection of this line
results a point node of SOP.

Comparing theoretical nodal structure of SOP with experimental one, we can
identify IRs corresponding to Cooper pairing [18,19]. Group theoretical
analysis of possible IRs in $k_{z}$ direction for $T_{h}$ group is carried
out in Table 1. It follows from this table, that for singlet pairs all even
one-dimensional IRs, but three dimensional IR is forbidden. For triplet
pairs without account for spin three dimensional only odd three dimensional
IR is possible, but all one-dimensional IRs are forbidden. When the spin is
included in the whole symmetry, all IRs are possible, but the type of IR
depends on the orientation of the spin relative to $Z$-axis. Triplet basis
functions for the case of $k_{z}$ directions are presented in Table 2. It
should by noted that despite of similarity \ of these functions to that of
ref. [28], physical significance is different. In refs.[1,12,13,28] basis
vectors\ $k_{x}$, $k_{y}$ and $k_{z}$ change continuously \ and one formula
(e.g. formula (12) of ref. [28]) describes all directions of $\vec{k}$. On
the other hand the data of Table 2 correspond to to the directions of
coordinate axes only. For $\vec{k}$ a general point in a Brillouin zone the
shape of our basis functions is different and we'll obtain it by projection
operator technique.

In general point of a Brillouin zone the dimension of the wave vector star
equals to the number of point group elements, and the number of basis
functions in each case (singlet or triplet) is two times less, since the
inversion was already used in the construction of functions (6) and (7).
Triplet basis function for IRs $A_{u}$ and $T_{u}$ constructed by projection
operator technique are \ presented in table 3. The nodal structure of basis
functions of $E_{1u}$ and $E_{2u}$ is the same as for $A_{u}$ and these
functions are not shown in this table. Function for $\vec{k}_{1}$ a general
point in a Brillouin zone is denoted as $\Phi _{1}^{t}$ and the results of
action on it by operations $h_{2}$, $h_{3}$ and $h_{4}$ [29] ( 180$^{o}$
rotations about axis X, Y and Z respectively) are denoted as $\Phi
_{2}^{t},\Phi _{3}^{t}$ and $\Phi _{4}^{t}$. The rotations $h_{5}$ and $%
h_{9} $ transform Z axis into positions Y and X respectively and function of
last row of IR $T_{u}$ into the second and the first row. There is single
spatial basis in the cases of IRs $A_{u}$, $E_{1u}$ and $E_{2u}$. When the
triplet spin in included the dimension of each basis is increasing in three
times. In the case of IR $T_{u}$ there are three independent spatial basis
sets, which can be obtained starting projections from different rows. These
basis sets are denoted in Table 3 by superscripts $\alpha $, $\beta $, and $%
\gamma $. It should be also taken into account that in each point in $\vec{k}
$- space there are three directions of spin are possible. Thus in the case
of IR $T_{u}$ we have 9 independent basis sets. This dimension follows from
the fact that total dimension of two electron basis equals to the square of
dimension of one-electron basis. We show only four of this basis sets $%
T_{u}^{\alpha }(\hat{x}),$ $T_{u}^{\alpha }(\hat{y})$, $T_{u}^{\alpha }(\hat{%
z})$ and $T_{u}^{\beta }(\hat{z})$.If the wave vector $\vec{k}_{1}$ is
approaching one of the coordinate axis four vectors connected by $180^{o}$
about this axis and reflections in planes normal to other axis coordinate
axis merge and there are two possibilities: basis functions transform into
the function presented in Table 2 or cancel. The latter case corresponds to
a point node of the wave functions. The nodal directions also shown in Table
3. Different basis sets for each of IRs $A_{u}$ and $T_{u}$ vanish at
different directions. Since in the absence of external magnetic field all
basis sets are physically equivalent, there are no symmetry requirements for
vanishing of pair functions in the symmetry directions. This conclusion
generalizes the Blount theorem for the case of symmetry directions in which
the dimension of IR of wave vector group equals to unity.

The above conclusion implies that experimentally observed point nodes of SOP
of $PrOs_{4}Sb_{12}$ are connected with symmetry violations. Experimentally
observe magnetic fields in $PrOs_{4}Sb_{12}$ result in time-reversal
symmetry violations [10]. In the presence of magnetic fields total spin
projections $M_{s}$=1 and $M_{s}$=-1 are not equivalent to each other and
its impossible to construct their real combinations $\hat{x}$ and $\hat{y}$.
In this case the triplet two-electron state splits into three states,
corresponding to tree components of total spin. According to Anderson [25]
the wavefunction of Cooper pair has a time -reversal and space inversion
symmetry. \ Singlet pair and $M_{s}$=0 pair (in our notations $\hat{z}$)
component of triplet invariant with respect to these operations.
Time-reversal interchanges two energetically non equivalent states $M_{s}$=1
and $M_{s}$=-1 and these state should be excluded from consideration. Also
spatial operations changing the direction of \ $\hat{z}$ component of spin
should be excluded. Thus in the case of time-reversal symmetry violation $%
\hat{z}$ component of triplet pair has a $D_{2h}$ symmetry.

Basis functions of $\hat{z}$ component of triplet pair in $D_{2h}$ symmetry
are presented in Table 4. In a general point of a Brillouin zone all IRs are
possible for such a pair, but different IRs reveal different structure of
point nodes at coordinate directions. Two-electron function of Au symmetry
has four nodes in directions $\ \pm X$ $\ $and $\pm Y$. Functions $B_{1u}$
and $B_{2u}$ have nodes in directions $\pm X$, $\pm Z$ and $\pm Y$, $\pm Z$
respectively. Since in the case under consideration Z is the direction of
magnetic field, these two functions are physically equivalent \ and only
nodes in directions $\pm Z$ \ follow from the symmetry. Finally, function of 
$B_{3u}$ symmetry has 6 nodes in directions of all coordinate axis. Hence,
there are tree possible nodal structures of SOP: Au with four nodes, $B_{1u}$%
+$B_{2u}$ with two nodes and $B_{3u}$ with six nodes.

\section{Discussion of the results}

\bigskip The phenomenological model $p+h$ function was proposed in ref.[30]
for $A$-pase in $PrOs_{4}Sb_{12}$ :

\begin{equation}
\Delta _{A}(k)=\Delta (1-k_{1}^{4}-k_{2}^{4}-k_{3}^{4})
\end{equation}

This formula describes six nodes proposed in ref. [8]. This nodal structure
corresponds to the $B_{3u}$ pair function of Table 4. For the $B-$ phase
with nodes in $Z-$ axis following formula was proposed in ref. [30]:

\begin{equation}
\Delta _{B}(k)=\Delta (1-k_{3}^{4})
\end{equation}

\bigskip This nodal structure corresponds to the sum of IRs $B_{1u}$ and $%
B_{2u}$ of Table 4. The model of of functions $s$- and $d-$ symmetry was
proposed in ref. [31]. This model describes the $A$- phase by the
anisotropic $s$- function with six nodes and $B$- phase. by the function $s$+%
$id_{x^{2}-y^{2}}$. In ref. [32] both states were described by the function $%
s$+$g$ type. Also the state with six node was described by the \ function of 
$f$- type [33]. Thus phenomenological approach enables to describe
experimental nodal structure of SOP by different types of functions. It
should be noted that making use of the reduction of basis functions of
rotational group onto the group $D_{2h}$, linear combinations of odd
spherical functions can be related to the functions of Table 4. Whereas,
since we used general rules for construction of two-electron wavefunctions,
our conclusions on te nodal structure of each IR don't depend on the basis
functions.

\section{Conclusion}

\bigskip Triplet two-electron wavefunctions for the space groups with point
groups $T_{h}$ and $D_{2h}$ are constructed. It is shown that total basis
for the group includes each one-dimensional IR three times and each three-
dimensional IR nine times and the basis functions of each IR vanish at
different directions. Thus in general case there are no symmetry reasons for
vanishing of SOP of any particular symmetry at any direction (Blount theorem
[13]). Spontaneous magnetic fields (time-reversal symmetry violations)
result in reduction of symmetry group of spin part to $D_{2h}$. In this case
there are two physically non-equivalent basis sets with zero projection of
total spin on quantization axis : basis of IR Au wit four nodes in
directions $\pm X$ and $\pm Y$, basis $B_{1u}+B_{2u}$ \ with two nodes in
directions $\pm Z$ and basis $B_{3u}$ \ with nodes in directions $\pm X,\pm
Y $ and $\pm Z$. Basis $A_{u}$ with four nodes (or basis $B_{3u}$ with six
nodes [8]) corresponds to $A$-phase of $PrOs_{4}Sb_{12}$ and basis \ $%
B_{1u}+B_{2u}$with two nodes corresponds to the $B-$phase. Thus group
theoretical account for the time-reversal symmetry violation made possible
to connect experimentally observed nodal structure of SOP in \ $%
PrOs_{4}Sb_{12}$\ \ with the index of IR of the symmetry group without
additional assumptions on the shape of basis functions.

\section{References}

1. M. Sigrist, K. Ueda. Rev. Mod. Phys. \textbf{63}, 239 (1991).

2. R.H. Heffner, J.L. Smith, J.O.Willis. et.al.Phys. Rev. Lett. \textbf{65}
, 2816 (1990).

3. K. Ishida, H. Mukuda, Y. Kitaoka, et.al. Nature. \textbf{396}, 658 (1998).

4. G.M.Luke, Y. Fudamoto, K.M. Kojima, et.al. Nature.\textbf{\ 394}, 558
(1998).

5. K.Maki, S.Haas, D.Parker, H.Won, K.Izawa, Y.Matsuda, Europhys.Lett. 
\textbf{68}, 720 (2004)

6. Y.Aoki, A. Tsuchiya, T.Kanayama, et.al. Phys. Rev.Lett. \textbf{91},
0067003 (2003).

7. E.E. Chia, M.B. Salamon, H. Sugawara, H.Sato. Phys.Rev.Lett. \textbf{91},
247003 (2002).

8. K.Izawa, Y,Nakajima, J. Goryo, et.al. Phys.Rev.Lett. \textbf{90}, 117001
(2003)

9. R.Vollmer, A.Fai$\beta $t, C. Pfleiderer. Phys.Rev.Lett. \textbf{90},
057001 (3003)

10. L. Shu, M.E. MacLaughlin, R.H.Heffner. et.al. Physica B.\textbf{\ 374},
247 (2006).

11. Y. Matsuda, K.Izawa, I,Vekhter. J. Phys. Cond. Matter. \textbf{18}, R705
(2006).

12. G.E. Volovik and L.P. Gor'kov: Sov. Phys. Jetp.\textbf{\ 61} , 843,
(1985).

13. E.I. Blount: Phys. Rev. B \textbf{32} , 2935, (1985).

14. S.Yip, A. Garg. Phys. Rev. B. \textbf{48}, 3304 (1993).

15. S.C. Zhang. Science, \textbf{275}, 1089 (1997).

16. V.G. Yarzhemsky and E.N. Murav'ev: J. Phys: Cond. Matter 4 , 3525 (1992).

17. V.G. Yarzhemsky: Zeitsch. Phys. B \textbf{99}, 19 (1995)

18. V.G.Yarzhemsky: Phys. stat. sol. (b) \textbf{209} , 101(1998).

19. V.G. Yarzhemsky: Int. J. Quant. Chem. Int. J. Quant. Chem. \textbf{80},
133 (2000).

20. V.G.Yarzhemsky , V.I. Nefedov. Int. J. Quant Chem.\textbf{\ 100}, 519
(2004).

21.V.G.Yarzhemsky , V.I. Nefedov. Doclady RAS. 404, 481 (2005)

22. V.G.Yarzhemsky , V.I. Nefedov. Phil. Mag. Lett., 86, 733 (2006).

23. V.G.Yarzhemsky , V.I. Nefedov. Doclady RAS. 412, 624 (2007) (in Russian)

24. V.L.Ginzburg, L.D.Landau, Zh.Exp.Teor.Fiz. \textbf{20},1064 (1950) (in
Russian)

25. P.W. Anderson: Phys. Rev. B \textbf{30 }(1984) 4000.

26. C.J. Bradley and A.P. Cracknell: \emph{The Mathematical Theory of
Symmetry in Solids. Representation Theory of Point Groups and Space Groups.}
Oxford, Clarendon, 1972.

27. R.A.Evarestov, V.P.Smirnov Methods of Group Theory in Quantum Chemistry
of Solids. S.Petersburs Univ. Press1987. (in Russian)

28. I.A. Sergienko, S.H. Curnoe, Phys. Rev. B. \textbf{70}, 144522 (2004).

29. O.V. Kovalev:\emph{\ Irreducible and Induced Representations and
Corepresentations of Fedorov Groups }Nauka 1986, (in Russian).

30. H. Won, Q. Yuan, P. Thalmeier, K. Maki. Braz. J. Phys.\textbf{\ 33}, 675
(2003).

31. J. Goryo, Phys.Rev.Lett.B. \textbf{67}, 184511 (2003).

32. K.Maki, H.Won, P. Thalmeier, Q.Yuan, K.Izawa, Y.Matsuda. Europhys.Lett. 
\textbf{64}, 496 (2003)

33. M.Ichoka, N. Nakai, K.Machida, J.Phys. Soc.Jap. \textbf{72}, 1322 (2003).

34.K. Miyake, H.Kondo, H.Harima. J.Phys.: Cond. Matt. \textbf{15}, L275
(2003).

\bigskip

Table 1. Symmetry of two-electron wavefunctions for the direction $k_{z}$
(group $T_{h}$)

\begin{tabular}{|l|l|l|}
\hline
{\small Basis} & {\small IR of }$D_{2h}$ & {\small IR of }$T_{h}$ \\ \hline
${\small \Phi }_{z}^{s}$ & ${\small A}_{g}$ & $A_{1g}${\small , }$E_{1g}$%
{\small , }$E_{2g}$ \\ \hline
${\small \Phi }_{z}^{t}$ & ${\small B}_{1u}$ & ${\small T}_{u}$ \\ \hline
${\small \Phi }_{z}^{t}{\small \hat{x}}$ & ${\small B}_{2u}$ & ${\small T}%
_{u}$ \\ \hline
${\small \Phi }_{z}^{t}{\small \hat{y}}$ & ${\small B}_{3u}$ & ${\small T}%
_{u}$ \\ \hline
${\small \Phi }_{z}^{t}{\small \hat{z}}$ & ${\small A}_{u}$ & $A_{1u}$%
{\small , }$E_{1u}${\small , }$E_{2u}$ \\ \hline
\end{tabular}

\bigskip

Table 2. Triplet two-electron wavefunctions for the star $\left\{
k_{x},k_{y},k_{z}\right\} $ (group $T_{h})$

\begin{tabular}{|l|l|l|}
\hline
{\small IR of }$T_{h}$ & {\small Function} & {\small IR of }$D_{2h}$ \\ 
\hline
\multicolumn{1}{|l|}{} & ${\small \Phi }_{x}^{t}{\small \hat{y}}$ & ${\small %
B}_{1u}$ \\ \hline
\multicolumn{1}{|l|}{${\small T}_{u}^{\alpha }$} & ${\small \Phi }_{y}^{t}%
{\small \hat{z}}$ & ${\small B}_{3u}$ \\ \hline
\multicolumn{1}{|l|}{} & ${\small \Phi }_{z}^{t}{\small \hat{x}}$ & ${\small %
B}_{2u}$ \\ \hline
\multicolumn{1}{|l|}{} & ${\small \Phi }_{x}^{t}{\small \hat{z}}$ & ${\small %
B}_{2u}$ \\ \hline
\multicolumn{1}{|l|}{${\small T}_{u}^{\beta }$} & ${\small \Phi }_{y}^{t}%
{\small \hat{x}}$ & ${\small B}_{1u}$ \\ \hline
\multicolumn{1}{|l|}{} & ${\small \Phi }_{z}^{t}{\small \hat{y}}$ & ${\small %
B}_{3u}$ \\ \hline
${\small A}_{1u}$ & ${\small \ \Phi }_{z}^{t}{\small \hat{z}+\Phi }_{y}^{t}%
{\small \hat{y}+\Phi }_{x}^{t}{\small \hat{x}}$ & ${\small A}_{u}$ \\ \hline
${\small E}_{1u}$ & ${\small \ \Phi }_{z}^{t}{\small \hat{z}+\varepsilon
\Phi }_{y}^{t}{\small \hat{y}+\varepsilon }^{2}{\small \Phi }_{x}^{t}{\small 
\hat{x}}$ & ${\small A}_{u}$ \\ \hline
${\small E}_{2u}$ & ${\small \ \ \Phi }_{z}^{t}{\small \hat{z}+\varepsilon }%
^{2}{\small \Phi }_{y}^{t}{\small \hat{y}+\varepsilon \Phi }_{x}^{t}{\small 
\hat{x}}$ & ${\small A}_{u}$ \\ \hline
\end{tabular}

\bigskip

Table 3. Triplet two-electron wavefunctions for the group $T_{h}$.

\begin{tabular}{|l|l|l|}
\hline
{\small IR} & {\small Function }$^{\ast )}$ & {\small nodes} \\ \hline
${\small A}_{u}{\small (\hat{x})}$ & ${\small (\Phi }_{1}^{t}{\small +\Phi }%
_{2}^{t}{\small -\Phi }_{3}^{t}{\small -\Phi }_{4}^{t}{\small )\hat{x}+(\Phi 
}_{9}^{t}{\small +\Phi }_{11}^{t}{\small -\Phi }_{12}^{t}{\small -\Phi }%
_{10}^{t}{\small )\hat{y}+(\Phi }_{5}^{t}{\small +\Phi }_{8}^{t}{\small %
-\Phi }_{6}^{t}{\small -\Phi }_{7}^{t}{\small )\hat{z}}$ & ${\small Y,Z}$ \\ 
\hline
${\small A}_{u}{\small (\hat{y})}$ & ${\small (\Phi }_{1}^{t}{\small -\Phi }%
_{2}^{t}{\small +\Phi }_{3}^{t}{\small -\Phi }_{4}^{t}{\small )\hat{y}+(\Phi 
}_{9}^{t}{\small -\Phi }_{11}^{t}{\small +\Phi }_{12}^{t}{\small -\Phi }%
_{10}^{t}{\small )\hat{z}+(\Phi }_{5}^{t}{\small -\Phi }_{8}^{t}{\small %
+\Phi }_{6}^{t}{\small -\Phi }_{7}^{t}{\small )\hat{x}}$ & ${\small X,Z}$ \\ 
\hline
${\small A}_{u}{\small (\hat{z})}$ & ${\small (\Phi }_{1}^{t}{\small -\Phi }%
_{2}^{t}{\small -\Phi }_{3}^{t}{\small +\Phi }_{4}^{t}{\small )\hat{z}+(\Phi 
}_{9}^{t}{\small -\Phi }_{11}^{t}{\small +\Phi }_{12}^{t}{\small -\Phi }%
_{10}^{t}{\small )\hat{x}+(\Phi }_{5}^{t}{\small -\Phi }_{8}^{t}{\small %
+\Phi }_{6}^{t}{\small -\Phi }_{7}^{t}{\small )\hat{y}}$ &  \\ 
\cline{1-1}\hline
\multicolumn{1}{|l|}{} & ${\small (\Phi }_{1}^{t}{\small +\Phi }_{2}^{t}%
{\small +\Phi }_{3}^{t}{\small +\Phi }_{4}^{t}{\small )\hat{x}}$ &  \\ 
\cline{1-2}
\multicolumn{1}{|l|}{${\small T}_{u}^{\alpha }{\small (\hat{x})}$} & $%
{\small (\Phi }_{9}^{t}{\small +\Phi }_{11}^{t}{\small +\Phi }_{12}^{t}%
{\small +\Phi }_{10}^{t}{\small )\hat{y}}$ & ${\small X,Y,Z}$ \\ \cline{1-2}
\multicolumn{1}{|l|}{} & ${\small (\Phi }_{5}^{t}{\small +\Phi }_{8}^{t}%
{\small +\Phi }_{6}^{t}{\small +\Phi }_{7}^{t}{\small )\hat{z}}$ &  \\ 
\cline{1-1}\hline
\multicolumn{1}{|l|}{} & ${\small (\Phi }_{1}^{t}{\small -\Phi }_{2}^{t}%
{\small -\Phi }_{3}^{t}{\small +\Phi }_{4}^{t}{\small )\hat{y}}$ &  \\ 
\cline{2-2}
\multicolumn{1}{|l|}{${\small T}_{u}^{\alpha }{\small (\hat{y})}$} & $%
{\small (\Phi }_{9}^{t}{\small -\Phi }_{11}^{t}{\small -\Phi }_{12}^{t}%
{\small +\Phi }_{10}^{t}{\small )\hat{z}}$ & ${\small X,Y}$ \\ \cline{2-2}
\multicolumn{1}{|l|}{} & ${\small (\Phi }_{5}^{t}{\small -\Phi }_{8}^{t}%
{\small -\Phi }_{6}^{t}{\small +\Phi }_{7}^{t}{\small )\hat{x}}$ &  \\ 
\cline{1-1}\hline
\multicolumn{1}{|l|}{} & ${\small (\Phi }_{1}^{t}{\small -\Phi }_{2}^{t}%
{\small +\Phi }_{3}^{t}{\small -\Phi }_{4}^{t}{\small )\hat{z}}$ &  \\ 
\cline{2-2}
\multicolumn{1}{|l|}{${\small T}_{u}^{\alpha }{\small (\hat{z})}$} & $%
{\small (\Phi }_{9}^{t}{\small -\Phi }_{11}^{t}{\small +\Phi }_{12}^{t}%
{\small -\Phi }_{10}^{t}{\small )\hat{x}}$ & ${\small X,Z}$ \\ \cline{2-2}
\multicolumn{1}{|l|}{} & ${\small (\Phi }_{5}^{t}{\small -\Phi }_{8}^{t}%
{\small +\Phi }_{6}^{t}{\small -\Phi }_{7}^{t}{\small )\hat{y}}$ &  \\ 
\cline{1-1}\hline
\multicolumn{1}{|l|}{} & ${\small (\Phi }_{5}^{t}{\small +\Phi }_{8}^{t}%
{\small -\Phi }_{6}^{t}{\small -\Phi }_{7}^{t}{\small )\hat{y}}$ &  \\ 
\cline{2-2}
\multicolumn{1}{|l|}{${\small T}_{u}^{\beta }{\small (\hat{y})}$} & ${\small %
(\Phi }_{1}^{t}{\small +\Phi }_{2}^{t}{\small -\Phi }_{3}^{t}{\small -\Phi }%
_{4}^{t}{\small )\hat{z}}$ & ${\small Y,Z}$ \\ \cline{2-2}
\multicolumn{1}{|l|}{} & ${\small (\Phi }_{9}^{t}{\small +\Phi }_{11}^{t}%
{\small -\Phi }_{12}^{t}{\small -\Phi }_{10}^{t}{\small )\hat{x}}$ &  \\ 
\hline
\end{tabular}

\bigskip *) Function $\Phi _{1}^{t}$ is defined by formula (7), the
functions $\Phi _{i}^{t}$ are the results of action of rotations of group $%
T_{h}$ in Kovalev [32] notations.

\bigskip

Table 4. Triplet two-electron wavefunctions for group $D_{2h}$.

\begin{tabular}{|l|l|l|}
\hline
{\small IR }$D_{2h}$ & {\small Function} & {\small Nodes} \\ \hline
${\small A}_{1u}$ & ${\small (\Phi }_{1}^{t}{\small -\Phi }_{2}^{t}{\small %
-\Phi }_{3}^{t}{\small +\Phi }_{4}^{t}{\small )\hat{z}}$ & ${\small X,Y}$ \\ 
\hline
${\small B}_{1u}$ & ${\small (\Phi }_{1}^{t}{\small -\Phi }_{2}^{t}{\small %
+\Phi }_{3}^{t}{\small -\Phi }_{4}^{t}{\small )\hat{z}}$ & ${\small X,Z}$ \\ 
\hline
${\small B}_{2u}$ & ${\small (\Phi }_{1}^{t}{\small +\Phi }_{2}^{t}{\small %
-\Phi }_{3}^{t}{\small -\Phi }_{4}^{t}{\small )\hat{z}}$ & ${\small Y,Z}$ \\ 
\hline
${\small B}_{3u}$ & ${\small (\Phi }_{1}^{t}{\small +\Phi }_{2}^{t}{\small %
+\Phi }_{3}^{t}{\small +\Phi }_{4}^{t}{\small )\hat{z}}$ & ${\small X,Y,Z}$
\\ \hline
\end{tabular}

\end{document}